\documentclass[conference, 9pt]{IEEEtran}
\usepackage{caption}
\usepackage{graphicx}
\usepackage{subcaption}
\usepackage{xcolor}
\usepackage[english]{babel}

\newcommand{\rw}[1]{Radioweaves}

\usepackage[binary-units=true]{siunitx}
\AtBeginDocument{\DeclareSIUnit{\Wh}{Wh}}

\usepackage{glossaries}
\newacronym{pse}{PSE}{power sourcing equipment}
\newacronym{pd}{PD}{powered device}
\newacronym{usrp}{USRP}{universal software radio peripheral}
\newacronym{rpi}{RPi}{raspberry pi}
\newacronym{poe}{PoE}{power-over-Ethernet}
\newacronym{ptp}{PTP}{precision time protocol}
\newacronym{vlc}{VLC}{visible light communication}
\newacronym{vlp}{VLP}{visible light positioning}
\newacronym{tsn}{TSN}{time-sensitive networking}
\newacronym{sdr}{SDR}{software-defined radio}
\newacronym{los}{LoS}{line-of-sight}
\newacronym{pps}{PPS}{pulse per second}
\newacronym{gpclk}{GPCLK}{general purpose clock}
\newacronym{ros}{ROS}{robot operating system}
\newacronym{lco}{LCO}{lithium cobalt oxide}
\newacronym{tof}{ToF}{time-of-flight}
\newacronym{ntp}{NTP}{network time protocol}
\newacronym{soc}{SoC}{state of charge}
\newacronym{uhd}{UHD}{USRP hardware driver}
\newacronym{pcb}{PCB}{printed circuit board}
\newacronym{pll}{PLL}{phase-locked loop}
\newacronym{os}{OS}{operating system}
\newacronym{hat}{HAT}{hardware attached on top}
\newacronym{gpu}{GPU}{graphical processing unit}
\newacronym{cpu}{CPU}{central processing unit}
\newacronym{if}{IF}{intermediate frequency}
\newacronym{fpga}{FPGA}{field-programmable gate array}
\newacronym{rfic}{RFIC}{radio-frequency integrated circuit}
\newacronym{ssd}{SSD}{solid state drive}
\newacronym{vep}{VEP}{virtual edge platform}
\newacronym{wr}{WR}{white-rabbit}
\newacronym{gnss}{GNSS}{global navigation satellite system}
\newacronym{mems}{MEMS}{micro-electromechanical systems}
\newacronym{cf}{CF}{cell-free}
\newacronym{wpt}{WPT}{wireless-power transfer}
\newacronym{daq}{DAQ}{data acquisition system}
\newacronym{iot}{IoT}{Internet of Things}
\newacronym{dlc}{DLC}{Distributed Laser Charging}

\usepackage[backend=biber,style=ieee]{biblatex}
\AtBeginBibliography{\footnotesize}

\usepackage{url}

\usepackage[super]{nth}

\begin{document}
\title{Techtile -- Open 6G R\&D Testbed for  Communication, Positioning, Sensing, WPT and Federated Learning\vspace{-1cm}}%

%Techtile -- A 6G Radioweaves Testbed\\ Infrastructure and R\&D Opportunities
\author{\IEEEauthorblockN{Gilles Callebaut, Jarne Van Mulders, Geoffrey Ottoy, Daan Delabie, Bert Cox, Nobby Stevens, Liesbet Van der Perre}
\IEEEauthorblockA{\IEEEauthorblockA{
        KU Leuven, WaveCore, Department of Electrical Engineering (ESAT), Ghent Technology Campus\\
        B-9000 Ghent, Belgium\\
        gilles.callebaut@kuleuven.be
    }}}

% make the title area
\maketitle
%\gilles{check order authorlist}

% no keywords

\begin{abstract}
% \liesbet{in de titel overwegen 6G R\&D Infrastructure for communication, positioning, and sensing?}
New concepts for next-generation wireless systems are being developed. It is expected that these 6G and beyond systems will incorporate more than only communication, but also sensing, positioning, (deep) edge computing, and other services. The discussed measurement facility and approach, named Techtile, is an open, both in design and operation, and unique testbed to evaluate these newly envisioned systems. Techtile is a multi-functional and versatile testbed, providing fine-grained distributed resources for new communication, positioning and sensing technologies. The facility enables experimental research on hyper-connected interactive environments and validation of new algorithms and topologies. The backbone connects 140~resource units equipped with edge computing devices, \glspl{sdr}, sensors, and LED sources. By doing so, different network topologies and local-versus-central computing can be assessed. The introduced diversity of i) the technologies (e.g., RF, acoustics and light), ii) the distributed resources and iii) the interconnectivity allows exploring more degrees and new types of diversity, which can be investigated in this testbed. 
\end{abstract}
%\keywords{Testbed, 6G, Software-Defined Radio, Precision Time Protocol, Power-over-Ethernet, RadioWeaves}

\glsresetall

\section{Introduction -- From 5G to Radioweaves}
\begin{figure}[h!]
\centering
    \centering
    \includegraphics[width=0.8\linewidth,trim={3cm 10.5cm 13cm 2cm},clip]{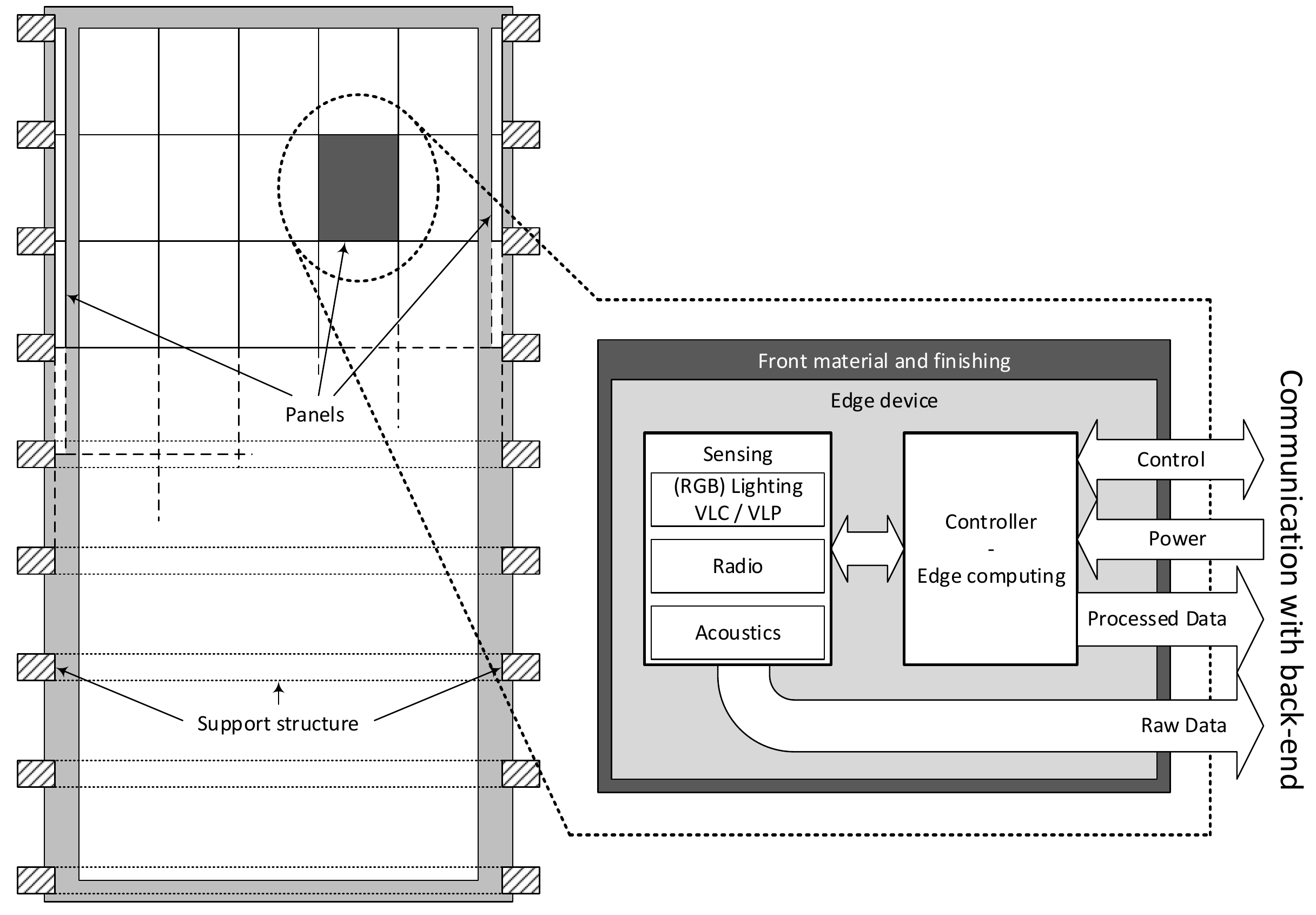}%
    \caption{\small Left: Support structure (top view) -- Right: Tile with embedded edge device. The support structure is designed to support 140~detachable tiles.}%
    \label{fig:System}
\end{figure}

%\bert{Estistisch, positioning van anker points, hoeveelheid data (sparse data), synchronizatie}
Massive MIMO~\cite{5595728}, used in 5G, has brought great improvements in both spectral and energy efficiency. By increasing the number of antennas at the base station, simple linear precoders (downlink) and detectors (uplink) could be utilized. Furthermore, the wireless channels between the users tend to become orthogonal, and the wireless medium behaves as if it were a flat fading channel. Conventionally, in massive MIMO, the base stations antennas are located together, often half-wavelength spaced, forming a large array. Further exploiting the spatial diversity, this approach has been extended to distributed massive MIMO, where several arrays are distributed over the coverage area. The extreme case of distributed massive MIMO is called \gls{cf} massive MIMO, where all antennas are distributed over the region and connected to a central processing unit, essentially removing the notion of cells. Spatially dispersed antennas are required to share data and frequency and timing information, making \gls{cf} networks not scalable in practice~\cite{9586055}.
To address these challenges, \rw has been introduced in~\cite{VanderPerreLiesbet2019Rfec}. There, it is envisioned that a `weave' of radios and other resources is created and integrated in existing structures, bringing them in closer to devices and eventually becoming truly ubiquitous. The Techtile infrastructure, illustrated in Figure~\ref{fig:System}, is a modular testbed capable of evaluating the \rw concept~\cite{VanderPerreLiesbet2019Rfec}. Such a system extends the conventional communication-only networks to provide other services, such as localization, sensing and \gls{wpt}.

\begin{figure*}[!htb]
      \centering
      \begin{subfigure}[t]{0.6\textwidth}
         \centering
         \includegraphics[height=2.2in]{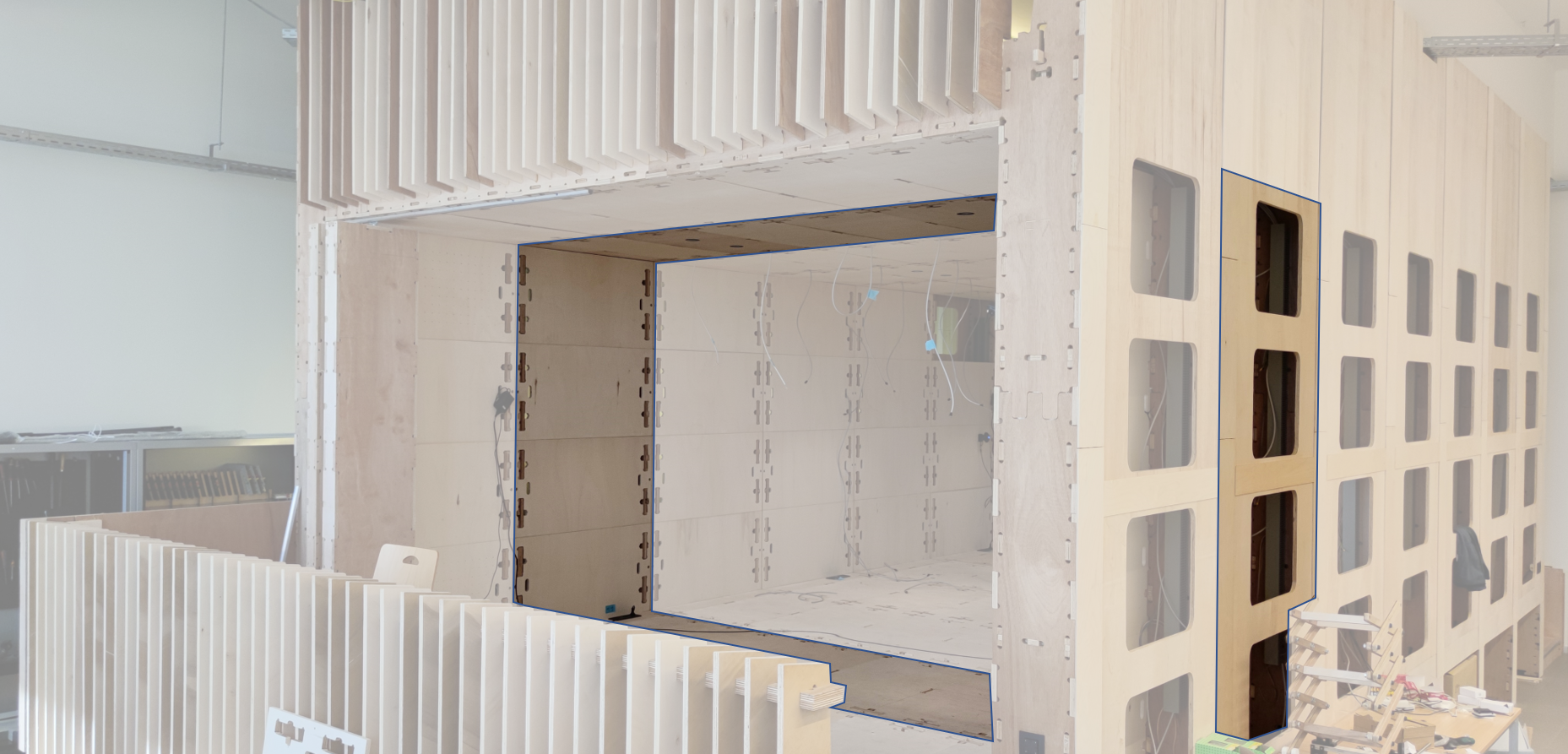}
         %\caption{{\small The Techtile infrastructure.}}\label{fig:techtile}
    \end{subfigure}%
    \qquad
    % \begin{subfigure}[t]{0.6\textwidth}
    %      \centering
    %      \includegraphics[width=\textwidth]{img/tile.jpg}
    %      \caption{{\small A tile equipped with a software-defined radio (USRP B210), processing unit (Raspberry Pi 4), power supply with Power-over-Ethernet.}}\label{fig:default-setup}
    % \end{subfigure}
    \begin{subfigure}[t]{0.25\textwidth}
    \centering
    \includegraphics[height=2.2in]{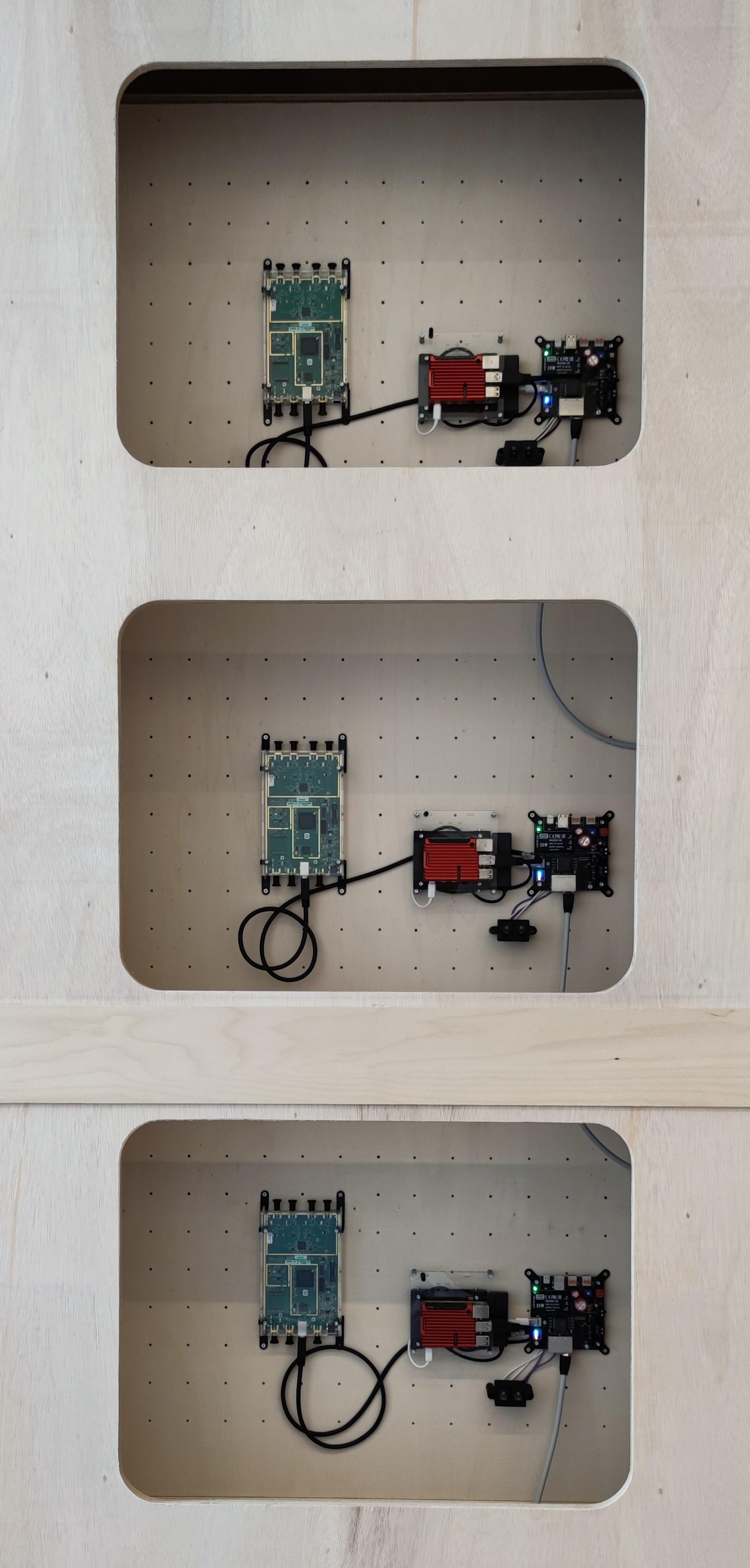}%
    %\caption{The back of three tiles, equipped with the default setup, i.e., a software-defined radio (USRP B210), processing unit (Raspberry Pi 4) and power supply with Power-over-Ethernet.}%
    %\label{fig:default-setup}
    \end{subfigure}
    \caption{\small Left: The Techtile support structure -- Right: The back of three tiles, equipped with the default setup, i.e., a software-defined radio (USRP B210), processing unit (Raspberry Pi 4) and power supply with Power-over-Ethernet. Each tile is connected to the central unit with an Ethernet cable, providing both power and data.}\label{fig:picture-techtile}
\end{figure*}

\begin{figure}[tbp]
\centering
    \centering
    \includegraphics[width=0.95\linewidth]{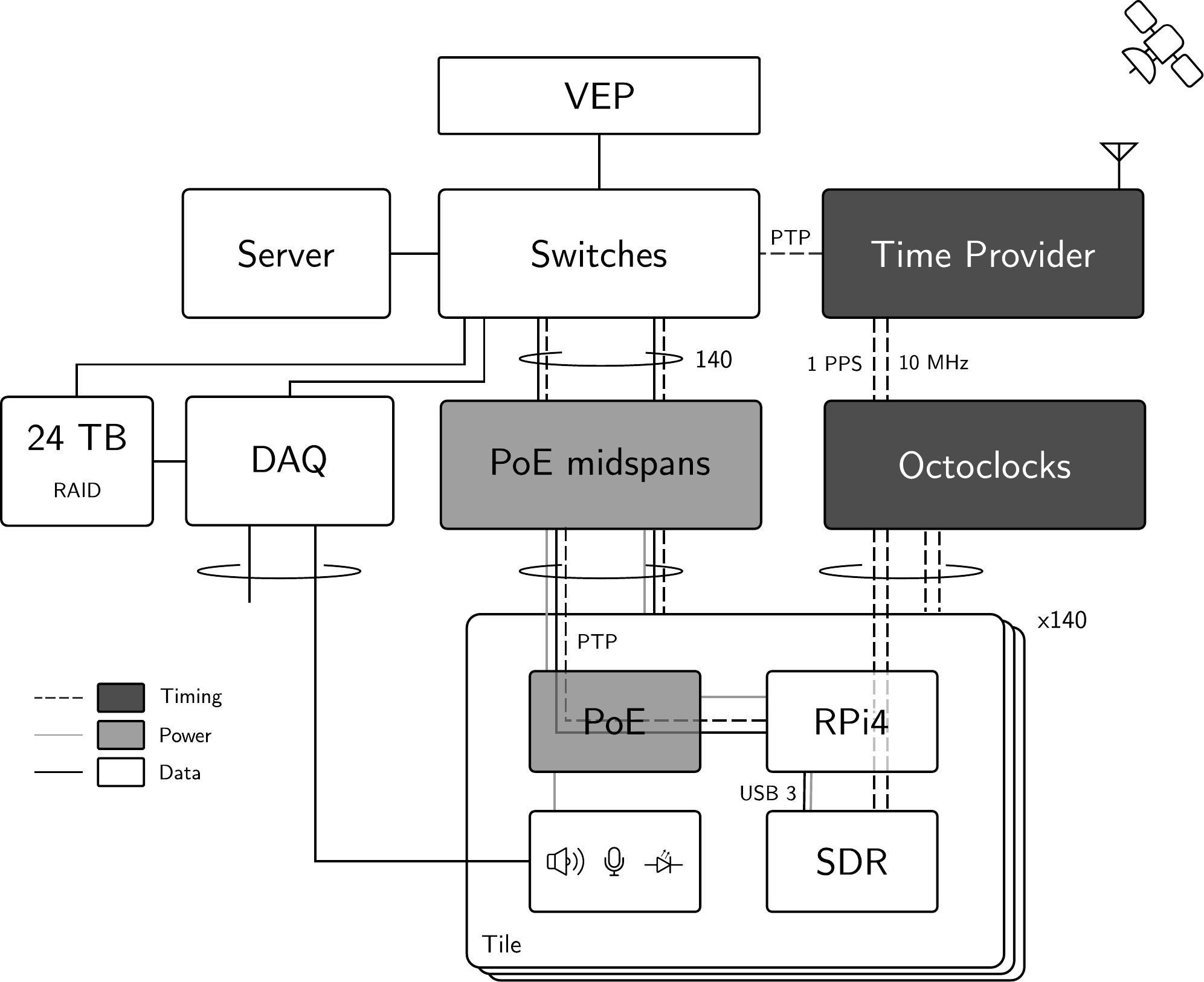}%
    \caption{\small Overview of the architecture of the testbed including logical links between components. Blocks highlighted in dark gray distribute synchronization information. Light gray blocks provision power through Ethernet or dedicated terminals and connectors. The solid lines depict connections exchanging data, e.g., Ethernet-based or USB. Dashed lines represent links transferring synchronization information.}%
    \label{fig:architecture}
\end{figure}

Main technical challenges, in both the testbed and future distributed infrastructures, are the scalability and synchronization.  In contrast to other testbeds~\cite{7063446,8471108,10.1145/3117811.3119863}, this R\&D infrastructure does not utilize dedicated connections for communication and synchronization to each processing point, i.e., --in our case-- tile. The conventional approach to time synchronization requires all cables to have the same length, making deployment cumbersome and not scalable. In order to support high-speed connections between the tiles and the central processing, multiple hierarchies of processing must be organized to aggregate the high number of separate connections. To tackle these issues, all tiles are connected, powered (PoE++ IEEE802.3bt) and time synchronized (\acrshort{ptp} IEEE 1588) over Ethernet. By default, each tile has a processing unit, a \gls{sdr} and a power supply, as depicted in Figure~\ref{fig:picture-techtile}. This base configuration can be extended with custom solutions to support other use cases. 
To facilitate other research activities, all developed equipment and software is open-source available.\footnote{\url{github.com/techtile-by-dramco}} An overview of the architecture of the Techtile testbed is depicted in Figure~\ref{fig:architecture}.

This paper is structured as follows. First, the open challenges of implementing distributed networking concepts in practical systems are elaborated, with a focus on the \rw concept. Next to this, the opportunities of having a spatially dispersed and diverse set of resources to provide different services is discussed in Section~\ref{sec:challenges-and-opportunities}. By first introducing the challenges and opportunities, the requirements imposed on the testbed become evident. In Section~\ref{sec:construction}, the modular construction is discussed. This is followed by the implementation of the back-end and on-tile resources in~\ref{sec:backbone} and \ref{sec:ontile}, respectively. Lastly, the envisioned research and development activities in the testbed are summarized in Section~\ref{sec:applications}, where we welcome, and even encourage, others to implement and evaluate their algorithms and methods in a practical scenario.

\section{Radioweaves -- Challenges and Opportunities}\label{sec:challenges-and-opportunities}

%\gilles{@liesbet, really your cup of tea. Could you expand/edit...?}

While the Radioweave concept and distributing compute-connectivity paradigms bears great potential~\cite{VanderPerreLiesbet2019Rfec}, key challenges need to be addressed in order to progress it towards practically deployable systems. These challenges stem from the high number of diverse resources hosted in the infrastructure and the novel application requirements. Prior to elaborating on the design of Techtile, some challenges are highlighted to demonstrate how the testbed enables research and development strategies to design, assess and prototype solutions to these challenges.

\subsection{Challenges}

\textbf{Heavy and distributed processing.}
6G applications significantly raise current requirements and pose new needs for services~\cite{REINDEERD1.1}. These include the provision of position information and mapping of the environment, interaction with energy-neutral devices enabled by \gls{wpt}, support for federated learning, and perceived zero-latency and ultra-reliability for critical services. An infrastructure hosting a high number of distributed antennas or wireless entry points and computing resources has the potential to accommodate these. Yet, this can lead to an extremely high volume of data propagating in the network, becoming a key factor for bandwidth and energy. However, it is expected that the information is rather sparse, i.e., not all data coming from all the resources contribute equally to the performance of the system and not all sampled data contain information that can not be disregarded.  

Novel digital signaling processing algorithms, including AI approaches, are being developed to address the expectations for 6G. Hereby, not all data can -- or is desired to -- be processed in a central manner, e.g., because of the high data volume or because of latency constraints. New algorithms and distribution of processing approaches are considered in order to reduce the data volume and latency within the performance requirements of the applications.  Next to this, the energy consumption of signal processing will become increasingly important. In previous cellular networks, the power amplifiers were responsible for the highest share of energy consumption~\cite{6065681}, while in 5G systems, this has been more balanced by the digital signal processing. One could expect this contribution to further increase with the number of resources, e.g., antennas.

% %~\cite{9443568}
% hoeveelheid data te verwerken, veelal sparse, edge vs central vs alles in between.
%  Increased impact on digital signalling processing. In conventional cellular systems, the energy expenditure was mainly due to the power consumption of the pas. In 5G systems, this has shifted towards the digital signalling processing. One could expect that this contribution would further increase with the number of resources, e.g., antennas.
 
% \textbf{Initial Access.} Due to the diverse set of wireless entry points into the infrastructure, the initial access is not trivial. 

\textbf{Synchronization and Calibration.}
In order to sample all signals coherently, all processing and sensing units need to be synchronized and calibrated. For instance, in order to exploit the reciprocity of the wireless channel, the hardware-dependent effects of the transmit and receive RF chains need to be measured~\cite{7880682}. After this calibration phase, the conventional uplink pilot-based method can be used for coherent transmitting and receiving when the  different units are synchronized in time (phase) and frequency. Due to the high number of dispersed resources in the infrastructure and the diverse services, synchronization, and calibration is an open challenge. Different approaches for synchronization and calibration can be evaluated in the testbed and is discussed in Section~\ref{sec:sync}.

%\gilles{angle of arrival with phase information, we need initial phase of RF chains. Part of the calibration phase.}

\textbf{Embedding in structures.}
To mitigate the impact on the environment and the users, the electronics will need to be fully embedded in existing structures, or they need to be integrated in a creative manner to enhance the user experience by, e.g., providing novel interaction possibilities with the technology. 
Fully embedding the electronics, is not feasible or desired when using technologies requiring a line-of-sight connection such as LEDs, microphones, or speakers. Next to the technological challenges, the perception of the visible technologies require careful design from an aesthetic point-of-view. 

\textbf{Organization of resources.}
An open research question is the spatial organization of individual resources to satisfy the requirements of the diverse applications. Radioweaves is expected to provide a wide range of services. For augmented reality applications, a dense network of resources close to the users could provide low-latency response and high-throughput wireless access. On the other hand, for positioning, geographically distributed beacons ensure the necessary diversity to enable fine-grained localization. Next to the positioning of resources, how these resources need to interconnected will be highly application-dependent.

\subsection{Opportunities}

\textbf{Distributed computing and storage resources.}
The availability of distributed computing and storage resources can improve on conventional wireless systems performance vs. complexity. For example, data reduction can be implemented near to the devices in monitoring systems, or short reaction times can be achieved for autonomous applications.
Interestingly, the distributed resources also open opportunities to support new services, such as federated learning~\cite{9141214}.

\textbf{Sensor/Technology Fusion.}
The multitude of different services, e.g., \acrlong{wpt} and communication, could be jointly exploited to enhance one another. For example, position information could improve the channel estimation procedure of the wireless channel. When we can estimate where a user will be, we can predict when high channel fading can be expected. 
In addition, different technologies providing the same service such as positioning with RF, acoustics and visible light can be combined to increase both the accuracy and the precision. 

\textbf{Increased Degrees of Freedom.}
The ample redundancy created in the points of access to the network is of particular importance to increase reliability. This avoids a single point of failure, and allows to anticipate on potential bad connections. Retransmission can be avoided, which is essential to achieve imperceivably latency.

\textbf{Spatial Diversity.}
The rich diversity can be exploited to improve positioning performance and especially achieve a high precision as the significant outliers, which are typical in highly reflective indoor environments, can be discovered and resolved.
The many simultaneous connections also increase the possibility to discriminate simultaneous streams.

\textbf{Close Proximity Operation.}
The dispersed access resources in the RadioWeave decrease the average distance of a device to the nearest point of entry to the network, and creates more favorable conditions with respect to antenna array angles. This leads to the possibility to reduce the transmit power of devices, and to achieve a higher energy budget when charging energy-neutral devices. Moreover, the proximity opens opportunities to reduce interference in the network.

\section{Techtile Construction -- Modular and Open}\label{sec:construction}
Figure~\ref{fig:picture-techtile} illustrates the Techtile support structure. It is based on the open-source building concept WikiHouse\footnote{\url{www.wikihouse.cc}}. Such constructions are made of standardized wooden parts. Building further on the modular design, the walls, floor and ceiling of the structure are comprised of 140~detachable tiles. The two walls of the building support 28~tiles each. The ceiling and ground floor support 42 and 52 tiles, respectively. The room's dimension is \SI{4}{\meter} by \SI{8}{\meter}, with a height of \SI{2.4}{\meter}. A tile has a size of \SI{120}{\centi\meter} by \SI{60}{\centi\meter}. To mount the electronics, the tile features a \SI{5}{\centi\meter} by \SI{5}{\centi\meter} grid of M3 inserts. This allows conveniently installing components with standardized M3 screws. In addition to attaching equipment and custom electronics to the tile, we allow designing your own tile for specific applications. For instance, \gls{vlc} and \gls{vlp} systems using LED fixtures can permanently be mounted in the ceiling. We also support attaching equipment to the inside of the room by means of overlays. An overlay has the same grid structure as a tile and is mounted on top of the tile. This keeps the aesthetics, while still supporting attaching equipment on the inside of the room. %As the infrastructure is mainly designed for embedding electronics inside walls, we emphasize on only using the overlays if absolutely required. 
This is required for technologies demanding a \gls{los}, e.g., visible light and acoustics. %Overlays are more flexible compared to designing your own tile, as overlays can be installed on all tiles.

% \begin{figure}
%     \centering
%     \def\svgwidth{\linewidth}\fontsize{8pt}{10pt}\selectfont
%      \input{img/plan.pdf_tex}
%     %\includegraphics[width=\linewidth]{img/plan.pdf}
%     \caption{\small 2 walls with 28 tiles each, a ceiling and the ground floor with 42 tiles  (excl. 10 optional tiles), 140 tiles. Each ring has its own color, see Fig.~\ref{fig:picture-techtile} where such a ring is highlighted. \gilles{nummering...}}
%     \label{fig:plan}
% \end{figure}

%\FloatBarrier
\section{Techtile Backbone -- Everything over Ethernet}\label{sec:backbone}
The backbone of the infrastructure consists of a central server, connecting all tiles via Ethernet. \SI{90}{\watt} of power is supplied to each tile by means of \gls{poe} midspans. The Ethernet switches support the IEEE-1588 \gls{ptp}, enabling high-accuracy clock distribution to all connected devices. 
The testbed, thus, provides communication, synchronization, and power over Ethernet. As everything is connected over Ethernet, the developed system is easily scalable and is flexible in the manner in which devices are added and removed from the network. Furthermore, by connecting all tiles using Ethernet, all different network topologies can be emulated, e.g., tree or mesh. This facilitates investigating the impact of different network topologies. 

\subsection{Central Processing and Networking}
The central server (Dell PowerEdge R7525) has \SI{512}{\giga\byte} RAM, two NVIDIA Tesla T4 \SI{16}{\giga\byte} \glspl{gpu} and two AMD 7302 \SI{3}{\giga\hertz} \glspl{cpu}, running Ubuntu Server 20.04 LTS.  At the networking side, four Dell S4148T-ON switches are used, featuring each 48 \SI{10}{\giga\bit} Ethernet ports with IEEE 1588v2 support. A Dell \gls{vep} 1485, running VyOS, handles routing, DHCP, VPN and firewall.

\subsection{Power-over-Ethernet}
\Gls{poe} technology, where both data and power go over the same Ethernet cable, is used to keep the cable management practical.  Our \gls{poe} architecture consists of several \gls{poe} midspans\footnote{The PD-96XXGC series midspans from Microchip is used.} supporting 156~\glspl{pd} (\gls{poe} clients).  Each midspan has a 10/100/1000 Mbps data rate pass through. In total, a \gls{poe} budget of approximately \SI{9}{kW} is available. 

% With the introduction of the latest standard, IEEE 802.3bt, a power of maximum \SI{90}{\watt} can be provided.
% While this maximum is not required for the default setup, implementing the latest standard ensures a generic solution, supporting also high-power applications. Our \gls{poe} architecture consists of several \gls{poe} midspans\footnote{The PD-96XXGC series midspans from Microchip is used.} supporting 156~\glspl{pd} (\gls{poe} clients).  Each midspan has a 10/100/1000 Mbps data rate pass through. In total, a \gls{poe} budget of approximately \SI{9}{kW} is available. 

\subsection{Synchronization}\label{sec:sync}
Multiple manners to distribute synchronization are available in the testbed to investigate different levels of synchronization, i.e., Ethernet-based, dedicated cabling or over-the-air synchronization.

\textbf{Over Ethernet.}
By default all tiles are time synchronized with \gls{ptp}~IEEE~1588v2~\cite{}. This protocol achieves a clock accuracy in the sub-microsecond range --and depending on the network, configuration, and version even the sub-nanosecond range. The protocol incorporates a master-slave architecture. The root time reference is hold by the grandmaster clock\footnote{Microchip TimeProvider 4100.} and is distributed to the other clocks in the network. \Acrlong{ptp} supports both L2 and UDP transport. It has an operation similar to \gls{ntp}, where the master and slave exchange messages to determine the path delays and correct their clocks accordingly. In \gls{ptp}, different profiles are defined, each having different configurations and requirements. Such profiles are tailored for specific application and are, for example, used in \gls{tsn}~\cite{8412458}.  Different clocks are defined based on their capabilities. For instance, transparent clocks are network devices that alter the timestamps in the packets to remove the time spent in these devices, effectively making them transparent to the \gls{ptp} protocol. A boundary clock, on the other hand, serves as both master and slave. It synchronizes its internal clock to a master on a specific port, and serves as a master on other ports. 

While the IEEE~1588v2 method ensures a scalable and low-cost solution, several other synchronization technologies exist. Recently, another Ethernet-based protocol, \gls{wr}~\cite{serrano2013white}, is being integrated in the IEEE~1588 protocol (IEEE~1588-2019)~\cite{9120376}. It extends the conventional IEEE~1588v2 protocol by including clock syntonization (frequency synchronization), phase detection (increasing the timestamp accuracy) and link asymmetry detection. These features result in sub-nanosecond accuracy time synchronization. At the time of writing, all required \gls{wr} switches are equipped with fiber optics and dedicated hardware \gls{wr} end-nodes are necessary to support \gls{wr}, making this solution not scalable in terms of costs and wiring for 140~end-device. 

\textbf{With dedicated cabling.}
To provide a reference for time and frequency synchronization, clock distribution modules\footnote{National Instruments OctoClock CDA-2990} can be used in the infrastructure. It provides a frequency and time reference via a \SI{10}{\mega\hertz} and 1\,\gls{pps} source, respectively. The clock distribution modules are synchronized via the grand master clock, which is in turn synchronized via \gls{gnss}. While this approach is not scalable, it serves as a baseline to compare the deterioration of the system services when the synchronization accuracy decreases or when high-accurate synchronization is required. % or when coherent operation of \glspl{sdr} is required without training.

% One of the ongoing work is to tune the configuration of the protocol to achieve nanosecond accuracy and determine the impact on the processing capabilities of the \gls{rpi} and the load on the network.

\textbf{Over-the-air.} Next to distributing synchronization information, the radio elements need to be calibrated to remove the hardware impairments and in particular mismatches between the many RX and TX chains. Radioweaves increases the technological difficulty to do this because of the high number of antenna elements and the ad-hoc nature of the infrastructure, i.e., there is no imposed design of connections and distribution of resources in space. Proposed over-the-air algorithms in literature, such as in~\cite{9569495,ganesan2021beamsync}, can be evaluated in this testbed.

%\FloatBarrier
\subsection{Data Acquisition System}\label{sec:DAQsection}
The \gls{daq}\footnote{The \gls{daq} consists of an 
18-Slot PXI Express chassis (NI PXIe-1095), peripheral modules and a controller (NI PXIe-8080 controller). The chassis holds 12 multifunction I/O modules (NI PXIe-6358), a timing and synchronization module (NI PXIe-6672) and a PXI bus extension module (NI PXIe-8394).} can provide the tiles with a synchronized analog data acquisition channel, able to sample at \SI{1.25}{MS/s}, with a resolution of 16~bit, and for a total number of 192~differential channels or 384~single-ended channels. 
This sample frequency and number of bits is sufficient to sample almost every potential sensor, which allows us to move the state-of-the-art in sensor fusion and underpins our aim for multi-modal sensing and positioning research. It includes sampling of full continuous audio streams (i.e., microphones recording audible or ultrasonic sound) and sensors registering high-bandwidth visible light communication and positioning~\cite{RaesWillem2022MLaa}. The high number of synchronized channels allows for setting up truly dispersed architectures, obtaining large and distributed sensor arrays. %Location bounding and sensitively monitoring the acoustic characteristics of the environment are key aspects for the research on physical layer security for embedded devices. \liesbet{komt een beetje uit de luch vallen, verwijzing naar verder opnemen?}

The \gls{daq} also has 48~synchronized 16-bit DAC output channels, with a sample rate of \SI{3.3}{MS/s},  able to steer a variety of actuators. For example, an array of speakers, ranging into the deep ultrasonic range, e.g. \SI{45}{k\hertz}, can be implemented in research on transmit beamforming and backscattering for fully passive mobile devices.
Additional possibilities are the generation of modulated signals and multiple access schemes to drive power LEDs in \gls{vlc} applications~\cite{RaesWillem2022MLaa}.

The combination of the \gls{daq} and the edge devices forms the infrastructure that is required to validate new concepts based on distributed nodes. By using the fully synchronized sensing capabilities of the central data acquisition (including a \SI{24}{T\byte} RAID storage) and processing infrastructure, a baseline performance can be established. This baseline performance can act as a reference for more realistic scenario's where distributed edge devices collaborate in a loosely coupled (and varying) configuration.

\begin{figure}[tbp]
\centering
    \centering
    \includegraphics[width=0.5\linewidth]{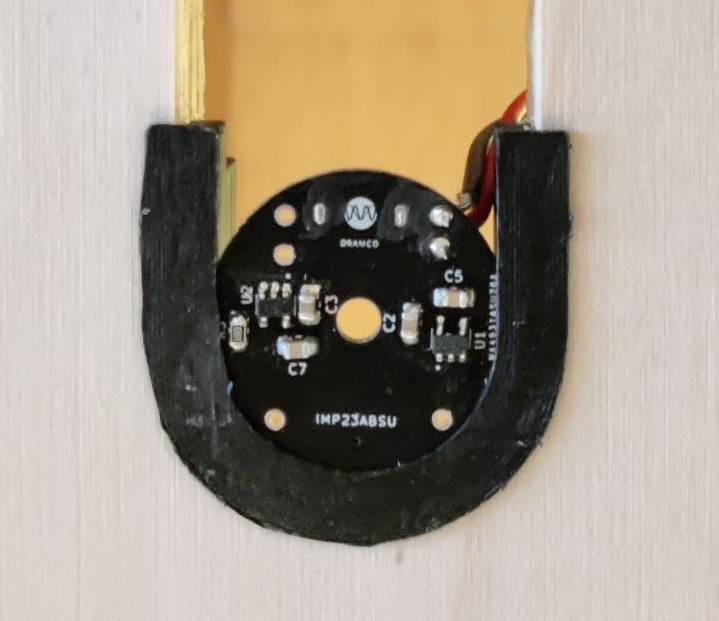}%
    \caption{A low-power \gls{mems} microphone with amplifier embedded in a tile.}%
    \label{fig:mic}
\end{figure}

\section{On-Tile -- Sensors, Radios, Processing, and Other Cool Stuff}\label{sec:ontile}
Each tile can accommodate a diversity of sensors and actuators, transmitters and receivers for radio - and other waves, host processing resources adding distributed intelligence in the environments, and potential other cool stuff that a creative researcher may want to experiment with.
In this section we elaborate on the \gls{poe} board, the RF communication through \glspl{sdr}, acoustic positioning, edge processing and automated 3D sampling via a rover.

% custom \gls{poe} board, the \gls{sdr}, the processing unit (\gls{rpi}) and . The \gls{poe} board delivers power to the tile and the \gls{rpi} processes the \gls{sdr} signals and serves as a platform for edge computing.

%\FloatBarrier
\subsection{PoE Board}

The standard \gls{poe} \gls{hat} for the \acrlong{rpi}, or other derivative forms,
are readily available but do not support the latest \gls{poe} version. For this reason, an 802.3bt supported \gls{pd} circuit was developed. %, as shown in Figure~\ref{fig:poe-board}. 
The system is based on the ON Semiconductor NCP1096 which is the PoE-PD interface controller. On top of the 802.3bt support, the board features several connectors and voltages to power different devices and sensors. %\Gls{poe} distributes \SI{48}{\volt} over Ethernet. 

\subsection{Wireless RF Communication through Software-Defined Radio}
The testbed  hosts a fabric of distributed \glspl{sdr}. Each tile is equipped with one \gls{usrp}~ NI~B210, featuring four RF channels (if in full-duplex mode). The B210 has a maximum transmit power of \SI{20}{dBm}.
This \gls{sdr} supports up to \SI{56}{\mega\hertz} of real-time bandwidth through the AD9361 direct-conversion transceiver. The B210 can operate over a frequency range of \SI{70}{\mega\hertz} to \SI{6}{\giga\hertz}, thereby covering most licensed and unlicensed bands. The B210 hosts an open and reprogrammable Spartan6 XC6SLX150 \gls{fpga}. The baseband signal is processed by the host, i.e., \gls{rpi}~4, using USB~3.0. GNURadio, supported by the B210, enables adopting and designing a high range of protocols and standards, e.g., IEEE 802.11 (Wi-Fi)~\cite{bloessl2013ieee} and LoRaWAN~\cite{9154273}, provided by the open-source community.
The B210 can be fed with an external \SI{10}{\mega\hertz} clock and \gls{pps} for synchronized operation (cfr.~Section~\ref{sec:sync}). The reference clock is used to generate all data clocks, sample clocks and local oscillators. In addition, an external \gls{pps} signal can be used for time synchronization between the \glspl{sdr}. In this manner, it is possible to coherently transmit and receive on all \glspl{sdr}, when properly calibrated~\cite{7356512} and/or trained~\cite{CallebautGilles2021AiSa}.

The baseband signals are processed by the \gls{rpi}~4. Depending on the application, the distributed signals are aggregated to the central server for further processing. By having both local processing and central processing, different techniques regarding local, edge, and cloud processing can be studied.

\subsection{Cool Other Stuff}

\textbf{Acoustics -- Microphones and Speakers.}
The Techtile infrastructure hosts both ultrasonic speakers and low-power acoustic \gls{mems} sensors to study acoustic positioning. The deployed \gls{mems} microphone is shown in Figure~\ref{fig:mic}. These can be sampled through the \gls{daq}, discussed in Section~\ref{sec:DAQsection}. Since our acoustic research focuses on frequencies between \SI{20}{\hertz} and \SI{45}{\kilo\hertz}, the current  sample rate of the \gls{daq} is amply sufficient. The offered 16-bit resolution is common for digitizing acoustic signals.

% \gilles{what needed for sampling acoustics signals? datarate, X-bit accuracy, ...}

% \gilles{first say what the infrastucre has to offer.}

\begin{figure}[tbp]
\centering
    \centering
    \includegraphics[width=0.5\linewidth]{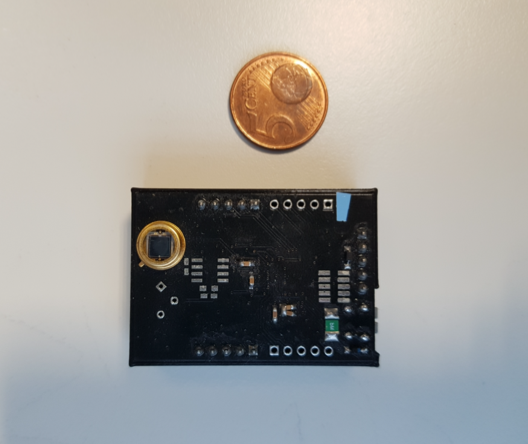}%
    \caption{Top view of the photo diode equipped receiver for optical wireless positioning.}%
    \label{fig:Rx_OWP}
\end{figure}

\textbf{Optical Wireless Positioning.}
As an emerging and high potential technology for indoor localization, optical wireless positioning is also integrated seamlessly in the infrastructure. LED sources, including the necessary drivers, are mounted in tiles in the ceiling. Figure~\ref{fig:Rx_OWP} shows the dedicated photo diode receiver equipped with a programmable system on chip (PSoC). As demonstrated in~\cite{RaesWillem2022MLaa},
highly accurate (P95 below 10 cm) and low latency (sub 100 ms) position estimates are established by means of embedded machine learning.

\subsection{Edge Processing}
The default setup contains a \gls{rpi} capable of edge computing. 
The latest \gls{rpi} model, i.e., \gls{rpi}~4, is adopted. It has a rich feature set, such as \SI{8}{\giga\byte} of LPDDR4-3200 SDRAM and a powerful processor, i.e., Quad core Cortex-A72 (ARM v8) 64-bit SoC 1.5GHz, tailored for high computing tasks.  Custom edge computing platforms can be used, e.g., NVIDIA Jetson~Nano, Google~Coral and Intel~NCS,  for dedicated and more computation-intensive applications. 

% To speed-up read/write operations, an \gls{ssd}, opposed to the standard micro-SD card storage, is used. Besides the speed improvement, the \gls{ssd} is less sensitive to power failure and has a higher capacity-to-price ratio. Furthermore, SD cards are designed to store data, while \glspl{ssd} are tailored for a high number of read/write operations, resulting from running an \gls{os}. These characteristics make \glspl{ssd} a more robust and suitable technology.

The combination of edge processing units and a central server allows researchers to evaluate different processing topologies and novel applications.

\subsection{Automated 3D Sampling}\label{sec:rover}

\begin{figure}[!htb]
    %  \hfill
    %   \begin{subfigure}[t]{0.48\textwidth}
    %      \centering
    %      \includegraphics[height=0.6\textwidth]{img/robot-base-plate.jpg}
    %      \caption{{\small Baseplate hosting the omnidirectional wheels, Controller (Raspberry Pi running \gls{ros}), battery pack and Marvelmind beacon.}}
    % \end{subfigure}
    % \hfill
    % \begin{subfigure}[t]{0.48\textwidth}
    %      \centering
    %      \includegraphics[height=0.6\textwidth]{img/robot-lift.jpg}
    %      \caption{{\small Scissor lift having a range of 55-\SI{185}{\centi\meter}.}}\label{fig:lift}
    % \end{subfigure}
    \centering
    \includegraphics[height=0.6\linewidth]{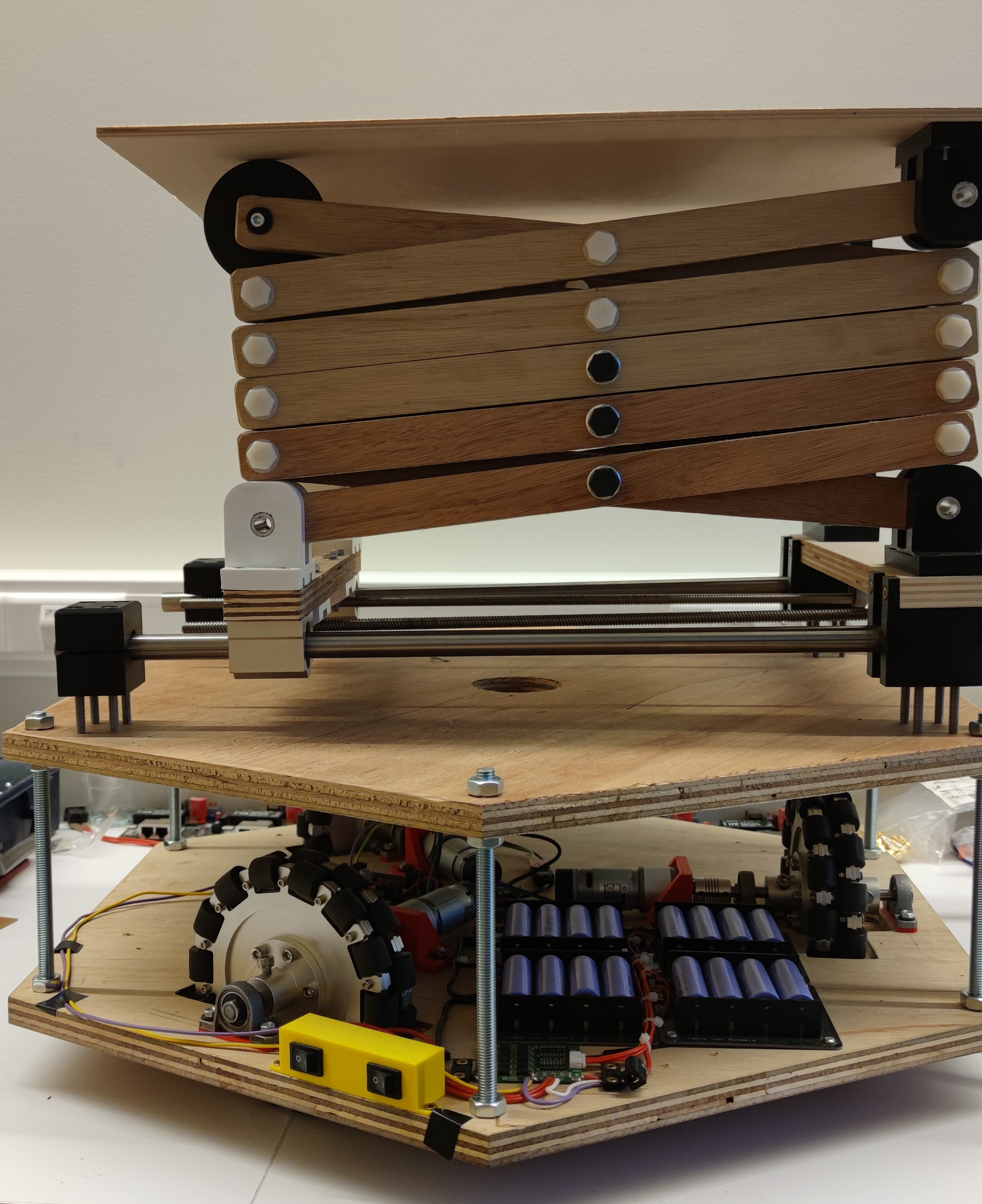}
    \caption{\small Rover to perform automated 3D sampling.}\label{fig:robot}
\end{figure}

To reduce labor-intensive tasks, speed-up measurements and mitigate human-errors, we have designed a rover to perform automated 3D sampling inside the testbed (Figure~\ref{fig:robot}). The rover consists of a baseplate, hosting the processing unit and batteries, and a scissor lift to move the measurement equipment to the required height.\footnote{This section contains work realized by two students, Arne Reyniers and Jonas De Schoenmacker, in the context of their master thesis.} Marvelmind indoor RTLS is used as a positioning system. It has a precision of \SI{2}{\centi\meter}~\cite{amsters2019evaluation}. Ultrasonic beacons are used to acquire the position of the mobile beacon (mounted on the baseplate). Four beacons are fixed in the Techtile infrastructure. The position is determined by trilateration. The rover features obstacle detection to mitigate crashing into objects in the room. Ultrasonic sensors mounted on the sides of the rover's baseplate are used and have a resolution of \SI{3}{\milli\meter} and a range of \SIrange{2}{400}{\centi\meter}. The navigation system can later be extended with other technologies, such as the one discussed in Section~\ref{sec:acoustic-positioning}. By including a scissor lift, the rover is able to measure in 3D space. The lift has a range of \SI{55}{\centi\meter} to \SI{185}{\centi\meter}. The measurement equipment can be mounted on top of the lift and powered via the on-board \SI{170}{\Wh} battery pack.

%%\FloatBarrier
\section{Research and Development Activities - Inviting Creative Experiments}\label{sec:applications}
The Techtile experimental facility is designed to enable a wide range of experiments. We foresee the following research and development activities: i) development of next-generation Internet-of-Things solutions, ii) experimental validation of beyond-5G communication, e.g., \rw, iii) positioning and sensing based on acoustic signals, iv) wireless charging and v) visible light communications and positioning. We conclude this paper with a glance on some ongoing research development.

\subsection{Precise Indoor Localization with Acoustics}\label{sec:acoustic-positioning}
% \gilles{move techtile-specific implmentation to previous section. This subsection is just to mention ongoing work specific to a radioweave setup.}
% Acoustic signals are a well adopted solution for indoor positioning \gilles{refs?}. Due to the low propagation speed (\SI{343}{m/s}), high-resolution clocks and high-speed microcontrollers can be omitted, enabling ultra-low power localization while maintaining centimeter accurate results. Notably, disadvantages, in contrast to RF-based systems, are the susceptibility to room acoustics (primarily reverberation) and the limited bandwidth of current low-power acoustic electronics. 

In previous work~\cite{CoxBert2021TCAa}, a hybrid RF-acoustic indoor positioning system was developed, exploring the possibilities of RF acting both as a synchronization and communication solution. This preliminary study shows that with a limited set of acoustic beacons and without any room information, a 3D~position can be estimated with a median Euclidean distance of \SI{0.75}{\meter} whilst consuming only \SI{362.45}{\micro\joule}. Further explorations in this research will make use of the Techtile infrastructure.
For example, the work of~\cite{CoxBert2021TCAa} will be extended to a large amount of hybrid RF-acoustic beacons. The available \glspl{sdr} are combined with 
ultrasonic speakers enabling ultra-low power RF communication in the form of backscattering.

% \begin{enumerate*}
%     \item hybrid RF-acoustic beacons in which the available SDR is combined with ultrasonic speakers enabling ultra-low power RF communication in the form of backscattering and wireless power transmission with electromagnetic waves
%     \item low-power acoustic MEMS sensors spatially embedded in the tiles, as depicted in Figure~\ref{fig:mic}, shifting the processing to the infrastructure side.
% \end{enumerate*}

% Thanks to the \gls{daq}, discussed in Section~\ref{sec:DAQsection}, fast prototyping is possible as all incoming and outgoing acoustic signals are effortlessly synchronized. Next to localization, 

% Thanks to the infrastructure's modular design, measurement configurations can be easily altered or extended. With Techtile, the possibilities for future, acoustic research are endless and we hope in the near future to perform acoustic Simultaneous Localization and Mapping (SLAM), sound classification for assisted living, %\bert{nog aanvullen met potentieel onderzoek}\daan{precise and reliable indoor localization for energy neutral devices, real-time object tracking and acoustic source localization ?}
%\gilles{@bert @daan}

\subsection{Positioning with Visible Light}
Our research~\cite{RaesWillem2022MLaa} has demonstrated that artificial neural networks (ANNs) and Gaussian Processes (GPs) provide an accuracy that is far superior to the results as obtained by a more classical multilateration process~\cite{9562197}. This especially in case nearby reflectors such as walls and obstacles are part of the setup, as will be evident in a \rw setup.

\subsection{Wireless Charging}
Next-generation smart homes and smart cities will host a massive amount of battery-powered or even battery-less devices, i.e., energy neutral devices. Different \acrlong{wpt} can be adopted, provisioning energy via electric, magnetic or electromagnetic fields. In order to support the diverse set of devices and applications targeted in Radioweaves, multiple \acrlongpl{wpt} need to be studied. For example, magnetic resonance coupling technology offers the advantage of being more flexible in charging devices and is a prime candidate for recharging robots and \gls{iot} nodes. To transfer energy wirelessly over larger distances,  technologies based on light~\cite{liu2016charging} and RF offer new opportunities. Energy neutral devices, without energy source included, could capture energy directly from RF power transmitters. In \rw, a high number of geographically dispersed RF transmitters are combined in the infrastructure, allowing to focus energy to the receiver by means of beamforming. As demonstrated in~\cite{7847396}, this approach is interesting for low-energy applications, e.g., smart light switches, electronic labeling, etc. Notably, this requires high accurate synchronization and calibration, as discussed in Section~\ref{sec:sync}. %Another approach~\cite{buyle2021multiband}, suggests using multiple frequency bands to overcome the transmit power restrictions imposed in unlicensed bands. Combining the beamforming and multiple band strategy becomes possible in this testbed due to the high number of transmit antennas (280). 

\section*{Acknowledgments}
This project has received funding from the European Union's Horizon 2020 research and innovation programme under grant agreement No. 101013425.

The realization of the Techtile infrastructure is made possible thanks to the generous support of a bequest granted to the Science, Engineering and Technology Group of the KU Leuven. We have received hardware from both Niko and On Semiconductor. We also thank our equipment suppliers National Instruments, Dell and W\"urth Electronics for their special support for this pioneering development.

{\footnotesize \printbibliography}

\end{document}